\documentclass[useAMS,usenatbib]{mn2e}
\usepackage{times}
\usepackage{graphics,epsfig}

\newcommand{\noh}{\ensuremath{N_{\rm OH}}}

\newcommand{\kms}{km~s$^{-1}$}
\newcommand{\cm}{cm$^{-2}$}

\newcommand{\noi}{\noindent}
\newcommand{\lb}{\left(}

\newcommand{\rb}{\right)}

\title[Detection of OH and wide HI absorption toward B0218+357]{Detection of OH and wide HI absorption toward B0218+357}

\author[Kanekar et al.]{N. Kanekar$^1$\thanks{E-mail: nissim@astro.rug.nl (NK);
chengalu@ncra.tifr.res.in (JNC); ger@nfra.nl (AGdB); dna@astro.tifr.res.in (DNA)},
J. N. Chengalur$^2$\footnotemark[1], A. G. de Bruyn$^{1,3}$\footnotemark[1], D. Narasimha$^4$\footnotemark[1] \\
$^{1}$ Kapteyn Institute, University of Groningen, Post Bag 800, 9700 AV Groningen, The Netherlands\\
$^{2}$ National Centre for Radio Astrophysics, Post Bag 3, Ganeshkhind, Pune 411 007, India \\
$^{3}$ Netherlands Foundation for Research in Astronomy, PO Box 2, 7990 AA Dwingeloo, 
The Netherlands \\
$^{4}$ Tata Institute of Fundamental Research, Homi Bhabha Road, Mumbai 400 005, India }

\begin{document}

\date{Received mmddyy/ accepted mmddyy}

\maketitle

\label{firstpage}

\begin{abstract}

We present deep GMRT OH and WSRT HI absorption spectra of the $z = 0.6846$ 
gravitational lens toward B0218+357. Both the 1665~MHz and 1667~MHz OH lines 
are clearly detected for the first time while a new wide absorption component 
was detected (at low significance) in the HI spectrum; the OH spectra yielded 
an OH column density of $N_{\rm OH} = 2.3 \times 10^{15}$~\cm. The ratio of 
1667 and 1665~MHz equivalent widths is $\sim 1.8$ while the redshift of peak 
OH absorption is $z = 0.68468 \pm 0.000008$ for both lines; this redshift agrees 
with that obtained from the HI line. The velocity spread (between nulls) of the HI 
absorption is $\sim 140$~\kms, while that of both OH lines is $\sim 100$~\kms; 
the HI and OH spectra are broadly similar in that they each have two principal 
narrow components and a wide absorption trough. We argue that the wide 
absorption is likely to arise from source components in the Einstein ring 
and derive a rotation velocity of $\sim 150$~\kms~at a distance of 1.5~kpc 
from the centre of the $z \sim 0.6846$ galaxy.
\end{abstract}


\maketitle

\section{Introduction}

	Gravitational lensing provides a new probe of the structure of high
redshift galaxies through their absorption lines. A typical spiral galaxy 
at $z \ga 0.5$ is usually too faint for detailed direct observations. 
However, when such an system gravitationally lenses a distant quasar, 
forming multiple images of the background source, spectra toward these 
images allow one to trace the kinematics along several lines of sight 
through the galaxy. Decimetre wavelength observations of such systems 
(i.e. spectra in the redshifted HI 21cm and OH radio lines) are particularly 
well suited to such studies since background sources typically show extended 
structure at these wavelengths, which allows a better sampling of the 
velocity field of the intervening galaxy. For example, Chengalur, de Bruyn 
\& Narasimha (1999) used Westerbork Synthesis Radio Telescope redshifted HI 
and OH spectra of the $z \sim 0.885$ lens toward PKS~1830-211 to estimate 
the rotation velocity of the absorbing galaxy. The relative strengths of 
the multiple OH radio lines (when detected) also allow one to trace the 
large scale distribution of molecular gas in the absorption system. However, 
the weakness of the OH transitions implies that they have only been detected 
in three absorbers at cosmological distance \citep{chengalur99,kanekar2002}.

One of the most well-known (and well-observed !) gravitational lenses
is the one at $z \sim 0.6846$ toward the source B0218+357 \citep{patnaik93}.
The radio continuum consists of two images of a flat spectrum source (components 
A and B), separated by 0.34$''$, and a radio ring with a radius of 0.18$''$ 
\citep{biggs2001}; this is the smallest known Einstein ring. The background
source is at a redshift of $z = 0.96$ \citep{lawrence96}; the high redshift of 
the lens and relatively low redshift of the source (with respect to the lens) 
imply that this system is an excellent candidate to determine the large scale 
geometry of the Universe and the value of the Hubble constant. In fact, VLA 
monitoring has yielded a time delay of $10.5 \pm 0.4$~days between the 
two images, resulting in a Hubble constant $H_0 = 69_{-19}^{+13}$~\kms~Mpc$^{-1}$ 
\citep{biggs99}. Unfortunately, however, Hubble Space Telescope (HST) observations 
have so far not been able to accurately locate the centre of the lens galaxy and 
this positional uncertainty implies a far higher uncertainty in $H_0$ than that 
derived from the statistical errors quoted above \citep{lehar2000,biggs2001}.

The $z = 0.6846$ system has been a rich source of absorption lines, with a number 
of molecular species such as CO, HCO$^+$, HCN, H$_2$CO and H$_2$O, as well as 
HI, already detected here \citep*{wiklind95,combes97,menten96,gerin97,carilli93}. 
Further, the difference between the rotation 
measures of components A and B has been measured to be $\sim 900$~rad~m$^{-2}$ 
and the optical spectrum of B0218+357 is highly reddened \citep{odea92,falco99}. 
All of these imply that the lens is an exceedingly gas-rich system; indeed, 
recent HST observations \citep{lehar2000} have shown that the 
system is a late-type spiral galaxy.

The original observations of the HI absorption profile \citep{carilli93} were
of low sensitivity and spectral resolution and hence could not resolve out the 
absorption feature. Further, the full width between nulls of the HI profile
was measured to be 75~\kms, somewhat small for a spiral galaxy at a moderate 
inclination, especially given the extended structure of the background continuum.
We present here deep Giant Metrewave Radio Telescope (GMRT) OH and Westerbork 
Synthesis Radio Telescope (WSRT) HI 
observations of the $z = 0.6846$ system; both the 1665~MHz and 1667~MHz OH 
lines were detected while strong limits were placed on the optical depth of the 
1720~MHz OH transition.
We also detect a wide component to the HI absorption, with a velocity spread 
of $\sim 140$~\kms. The observations and data analysis are discussed in 
section~\ref{sec:obs} while the implications for the $z = 0.6846$ absorber 
are discussed in section~\ref{sec:discussion}.

\section{Observations and data analysis}
\label{sec:obs}
\subsection{WSRT HI observations}
\label{sec:wsrt}

Observations of the redshifted HI line toward B0218+357 were carried out 
with the WSRT on 21~June 1998, with the new DZB correlator as the backend. 
A bandwidth of 2.5 MHz was used for the observations, centered at an observing 
frequency of 843.20~MHz and sub-divided into 256 channels. No taper was applied 
in the lag-to-frequency transform, yielding a velocity resolution of $\sim 
3.5$~\kms~(channel resolution $\sim 9.8$~kHz). 12 telescopes were used for the observations. 

The standard calibrators 3C48 (30min, before) and 3C286 (72min, after), were used to 
calibrate the absolute flux scale and the system bandpass; their adopted flux densities 
are on the Baars et al. (1977) scale. The total on-source time was 11.7 hours.

\subsection{GMRT OH observations}
\label{sec:gmrt}

The GMRT OH observations of B0218+357 were initially carried out on 
the 17th and 27th of June, 2001, with the standard 30 station FX correlator
as the backend. A total bandwidth of 4 MHz was used for these observations,
to include both the main OH transitions (at rest frequencies of 1665.4018~MHz
and 1667.3590~MHz). This was sub-divided into 128 channels, yielding a frequency
resolution of $\sim 31.25$~kHz (i.e. $\sim 9.4$~\kms) on each run. Twenty three 
and seventeen antennas were available for the observations on the 17th and the 
27th respectively, due to various debugging and maintenance activities. The standard 
calibrator 3C48 was used to carry out absolute flux and bandpass calibration. Total 
on-source times were 3 and 5 hours on the first and second observing dates
respectively.

While the 4~MHz bandwidth observations were important for the initial 
detection of the OH lines, it was apparent that their relatively poor velocity 
resolution would not allow the positions or widths of the absorption features to be 
determined as accurately as in the case of the HI line (note that the GMRT 
correlator allows a higher frequency resolution only at the cost of total 
bandwidth and the bandwidth of 4~MHz was necessary to include both OH 
features in the band). Further observations (each using a bandwidth of 1~MHz, 
i.e. a resolution of $\sim 7.8$~kHz or $\sim 2.4$~\kms) were hence separately carried 
out of the 1665~MHz and 1667~MHz features, on the 11th of October and the 27th of November, 
2001, respectively. 21 antennas were available for each of these observations; 
the total on-source times were $\sim 6$ hours and $\sim 3$ hours for the 1665~MHz 
and 1667~MHz observations respectively. Absolute flux and bandpass calibration 
were carried out using 3C48 and 3C147.

Finally, an attempt was also made with the GMRT to detect the 1720~MHz OH 
transition in the $z \sim 0.6846$ absorber. These observations were carried 
out on the 25th of November, 2001, using a bandwidth of 1~MHz (i.e. a 
resolution of $\sim 7.8$~kHz or $\sim 2.3$~\kms). Flux and bandpass calibration were again 
carried out using 3C48 and 3C147. The total on-source time was $\sim$
3 hours, with 28 antennas.

\subsection{Data analysis and spectra}
\label{sec:analysis}

\begin{figure}
\epsfig{file=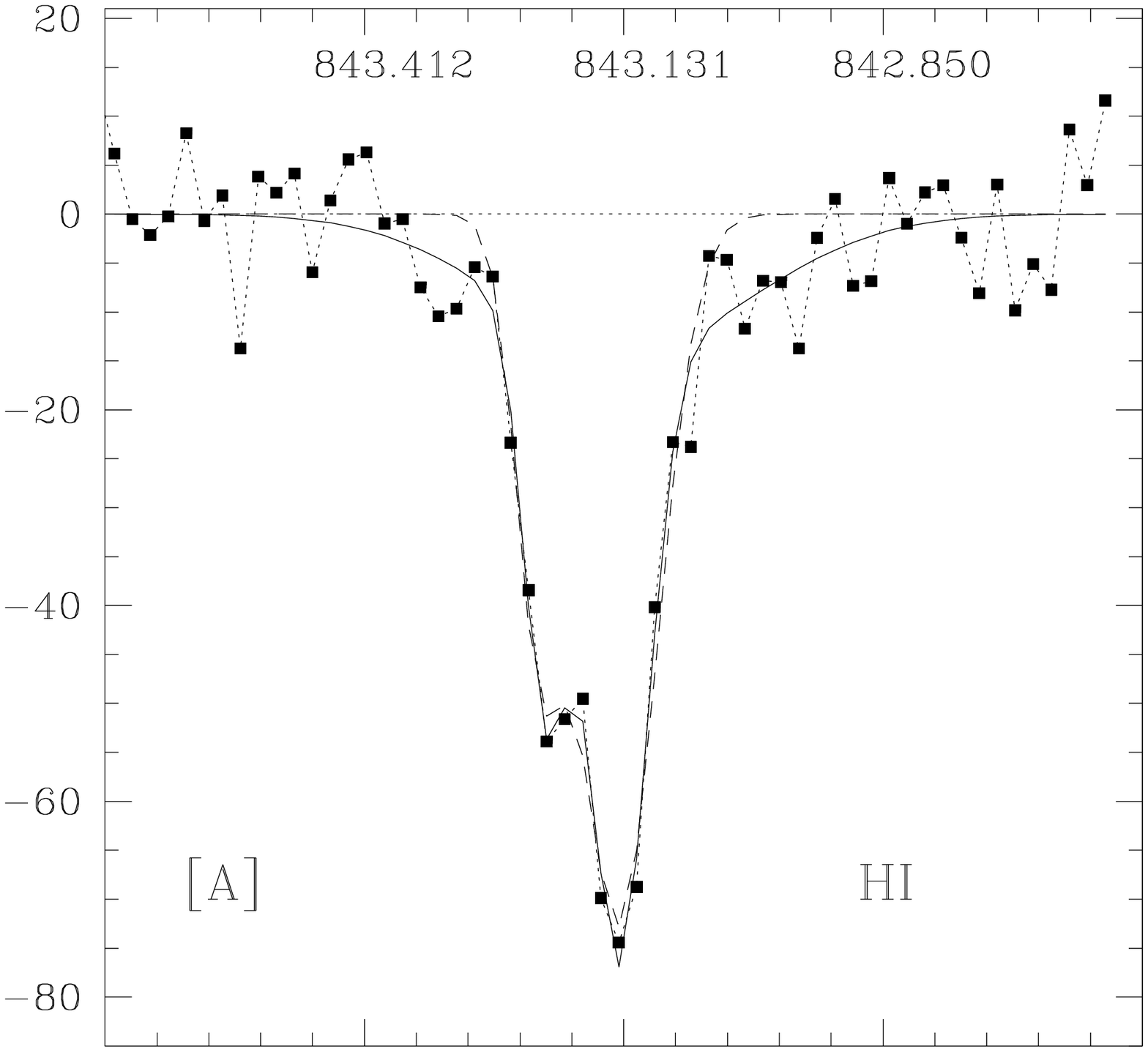,width=3.3in,height=3.2in}
\vskip -0.54in
\epsfig{file=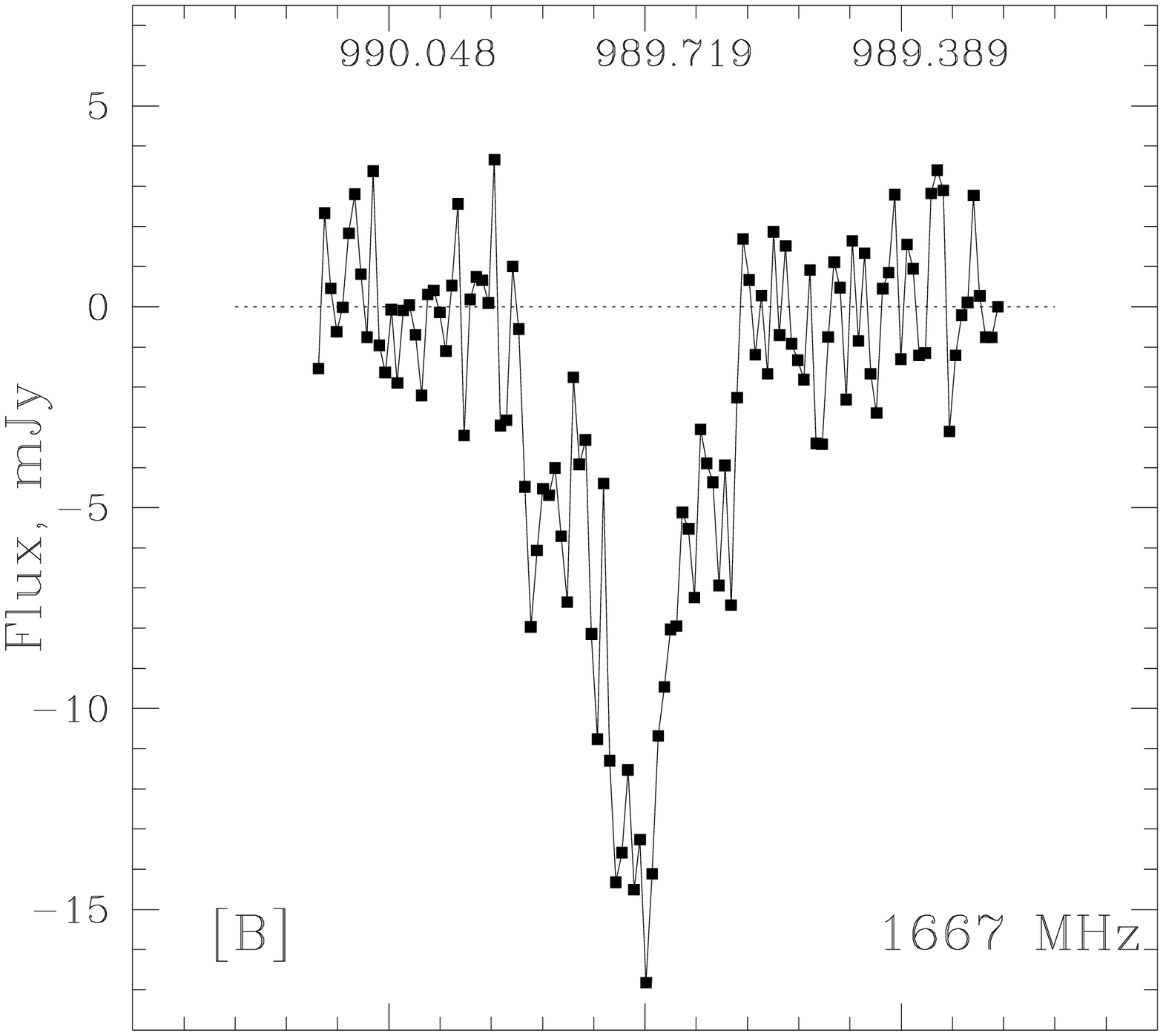,width=3.3in,height=3.2in}
\vskip -0.54in
\epsfig{file=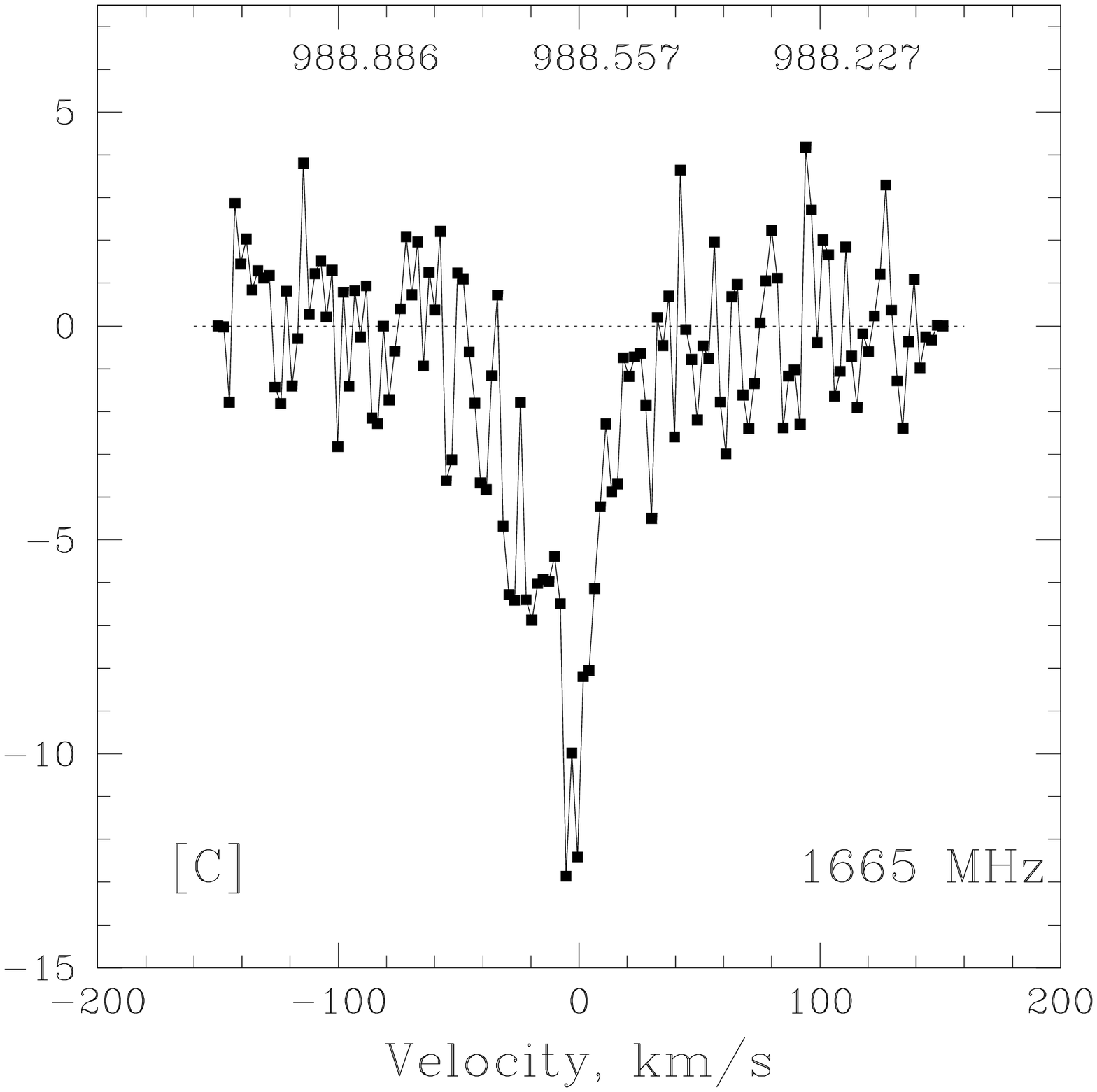,width=3.3in,height=3.2in}
\vskip -0.1in
\caption{ Absorption spectra toward B0218+357 : [A] WSRT 7~\kms~resolution HI 
	spectrum (solid squares); the solid line shows the 3-Gaussian fit. 
	[B] GMRT 2.4~\kms~resolution spectrum in the 1667~MHz OH line and 
	[C] GMRT 2.4~\kms~resolution spectrum in the 1665~MHz OH line. The x-axis 
	of each panel is velocity, in km/s, relative to $z = 0.68468$; the heliocentric 
	frequency scale (in MHz) for each line is at the top of each panel.
	}
\label{fig:spectra}
\end{figure}

The data were converted from the telescope format to FITS in all cases
and then analyzed in AIPS using standard procedures. A careful inspection
was initially carried out for RFI and sections showing evidence of interference
were flagged out. For the WSRT data, a simple average of the visibilities was
made after bandpass calibration and the peak flux density read off from
the image cube; this resulted in the spectrum shown in figure~\ref{fig:spectra}.
We note that B0218+357 is largely unresolved on WSRT baselines; the spectrum
represents the flux density averaged over 60 interferometers. A total of 6
baselines were not used in the averaging for various reasons. In the case of
the GMRT observations, after phase and bandpass calibration, the continuum emission
was subtracted by fitting a linear polynomial to the U-V visibilities, using the AIPS
task UVLIN. The continuum-subtracted data were then mapped in all channels and
spectra extracted at the quasar location from the resulting three-dimensional data
cube. In the case of the multi-epoch 4~MHz GMRT observations, the spectra of the
two epochs were corrected to the heliocentric frame outside AIPS and then averaged
together.

\begin{figure}
\centering
\epsfig{file=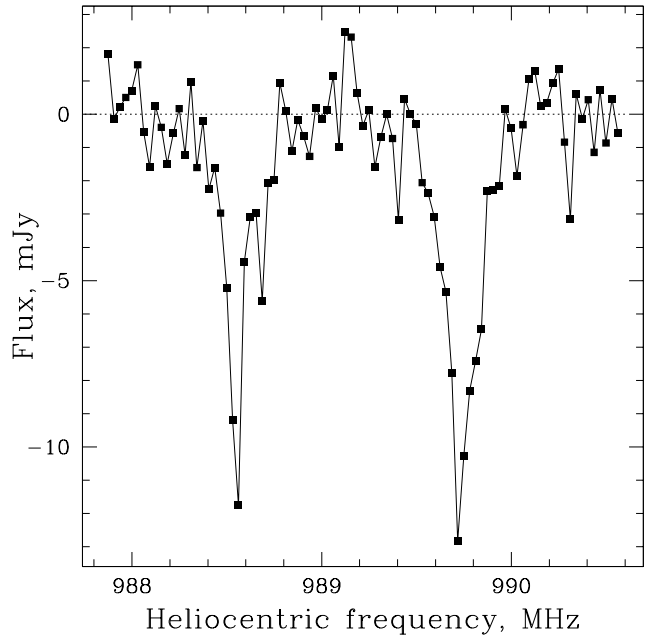,width=3.5in}
\caption{GMRT 9.4~\kms~resolution OH spectrum toward B0218+357. The 1665~MHz 
and 1667~MHz OH lines can be clearly seen.}
\label{fig:4MHz}
\end{figure}

The flux density of B0218+357 was measured to be 1.520~Jy (at 843~MHz) in the WSRT 
observations. In the case of the GMRT observations, the flux density was measured to 
be 1.64~Jy (17th and 27th of June), 1.54~Jy (11th of October) and 1.58~Jy (27th of 
November) at $\sim 989$~MHz 
and 1.66~Jy at $\sim 1021$~MHz (25th of November). Note that all spectra of the 
1665~MHz and 1667~MHz lines have been scaled here to a common flux density 
of 1.64~Jy. Earlier experience with the GMRT indicates that the flux calibration 
is reliable to $\sim 15$\%, in this observing mode.

The WSRT HI spectrum is shown in figure~\ref{fig:spectra}[A]; the spectrum has been 
Hanning smoothed and re-sampled and has a velocity resolution of $\sim 7$~\kms~and an RMS 
noise of $\sim 7$~mJy. Two deep features can be immediately identified in the 
spectrum, besides a weak broad component. The peak optical depth is 0.05, occuring 
at a redshift of $0.68467 \pm 0.00001$. The total velocity spread of the absorption 
is $\sim 150$~\kms. The spectrum has a total velocity coverage of $\sim 800$~\kms; 
only the central 400~\kms~are shown here, for comparison with the high resolution 
OH spectra. We note that non-Gaussian behavior was observed at two locations in the 
spectrum, near the edges of the band (beyond the velocity range plotted in 
figure~\ref{fig:spectra}~[A]); these are likely to be due to low-level RFI.

The final GMRT 4~MHz OH spectrum toward B0218+357 is shown in figure~\ref{fig:4MHz}.
The RMS noise is 1.1 mJy, per 9.4~\kms~channel; the spectrum has not been smoothed. 
Both OH transitions are clearly visible; the peak absorption occurs at 
heliocentric frequencies of 988.56 MHz and 989.72 MHz, both implying
a redshift $z = 0.68468 \pm 0.00003$. A wide shoulder can be seen in both 
absorption features, on the high frequency side of the principal feature.

Figures~\ref{fig:spectra}[B] and \ref{fig:spectra}[C] show the 1~MHz bandwidth 
GMRT spectra of the 1667~MHz and 1665~MHz OH lines. The RMS noise values are 
1.5~mJy and 1.6~mJy respectively, per 2.4~\kms~channel (the spectra have again not 
been smoothed).
The high frequency shoulder of the 4~MHz spectrum is now resolved out in both 
spectra; besides this, a wide absorption component can also be seen (which is 
stronger in the 1667~MHz transition). The equivalent width of the 1667~MHz 
absorption line is $\int \tau_{1667} {\mathrm d}V = 0.40 \pm 0.01$~\kms, while the 
ratio of the 1667~MHz and 1665~MHz equivalent widths is $R \sim 1.8$, as expected 
in optically thin conditions in thermal equilibrium. Interestingly, this ratio
is slightly larger at the low frequency edge of the two profiles (at $\sim 20$~\kms). 
The total velocity spread of both the 1665 and 1667~MHz lines is $\sim 100$~\kms. 

The peak optical depth in the 1~MHz bandwidth spectra occurs at heliocentric frequencies 
of 988.575~MHz (1665~MHz transition) and 989.717~MHz (1667~MHz transition), 
corresponding to redshifts $0.68465 \pm 0.000008$ and $0.68468 \pm 0.000008$ 
respectively. While the above redshifts are in marginal disagreement, 
figure~\ref{fig:spectra}[C] shows that the 1665~MHz line has two peaks of 
approximately the same depth, separated by $\sim 7.2$~\kms.  If the second 
(marginally weaker) feature is used to compute the redshift, one obtains 
$z_{1665} = 0.68468$, in agreement with the 1667~MHz redshift; the same value 
is also obtained by a Gaussian fit to the 1665~MHz line.  We will hence use this 
value for the redshift of the absorber, as it also agrees with that estimated 
from the 1667~MHz line. This is also in broad agreement with redshifts obtained 
from millimetre-wave transitions (e.g., $z = 0.684693$, from the $^{13}$CO line; 
\citealt{wiklind97}). Note that a measured difference between the redshifts 
of the four OH 18cm lines can, in principle, be used to simultaneously measure 
variations in the fine structure constant $\alpha$, the ratio of electron mass 
to proton mass $m_e/m_p$ and the proton g-factor $g_p$ \citep{chengalur03}.

Finally, the GMRT spectrum of the 1720~MHz OH transition (not shown here) 
did not show any absorption. The RMS noise was measured to be 1~mJy, 
after smoothing the spectrum to a resolution of $\sim  10$~\kms; this yields 
a $3 \sigma$ limit of 0.0018 on the optical depth in the 1720~MHz line, 
per 10~\kms.

\section{Discussion}
\label{sec:discussion}

The OH and HI profiles are broadly similar in that each absorption line consists
of a primary component, a high frequency shoulder and a broad shallow 
trough; this indicates that the lines originate in the same diffuse or 
dark cloud. While the total width of the HI absorption is wider than that of the 
OH lines ($\sim 150$~\kms~for the HI absorption against $\sim 100$~\kms~for the 
OH), this is not too surprising since OH is believed to be more confined than HI 
in models of molecular clouds (see, for example, figure~6 in \citealt{liszt96}). 
It is tempting to identify the two main features of the lines as stemming from 
absorption against the two point images A and B; however, VLBI observations of 
the HI absorption \citep{carilli2000} have shown that it arises solely due 
to image A, while no absorption is seen against image B. Similarly, the mm-wavelength 
molecular absorption lines are also only seen against image A and not against B 
\citep{menten96}. The two main features of the lines hence probably arise from 
the two VLBI components of image A (Patnaik et al. 2003, in preparation) or 
possibly from multiple clouds 
along the line of sight to one of these components, while the broad but shallow trough 
is likely to stem from absorption against source components in the radio ring.
 
For an optically thin cloud in thermal equilibrium, the OH column density of
the absorbing gas \noh~is related to the excitation temperature $T_x $ and the
1667 MHz optical depth $\tau_{1667}$ by the expression (e.g. \citealt{liszt96})
\begin{equation}
N_{\rm OH} = 2.24 \times 10^{14} {\lb {\frac {T_x }{f}} \rb}\int \tau_{1667}
 \mathrm{d} V \; ,
\label{eqn:noh}
\end{equation}
\noi where $f$ is the covering factor of the absorber. Here, \noh~ is
in cm$^{-2}$, $T_x $ in K and $\mathrm{d}V$~in \kms. The integrated optical depth 
in the 1667 MHz line is $\int \tau_{1667} \mathrm{d}V = 0.659$~\kms; thus, 
$\noh = 1.1 \times 10^{14} \times ({T_x / f})$~\cm.  VLBI observations 
\citep{carilli2000} have shown that no HI absorption is seen against component 
B; similarly, the CO and HCO$^+$ absorption are only seen against component A 
\citep{wiklind95}; the covering factor $f$ is thus likely to be close to 0.4.
Next, the OH excitation temperature $T_x $ cannot be directly estimated for cosmologically 
distant objects such as the $z \sim 0.6846$ absorber. In the Galaxy, OH emission 
studies have shown that this temperature may be as low as $T_x \sim T_{CMB} + 1$~K, 
with similar values for the HCO$^+$ line ($T_x \, (HCO^+) \sim T_{CMB}$; \citealt{lucas96}). 
However, the excitation temperatures of the redshifted HCO$^+$ lines in the four 
known high redshift absorbers have been found to be {\it higher} than 
$T_{CMB}(1 + z)$, the redshifted CMB temperature; it is thus quite likely that 
the OH excitation temperature too is higher in these systems. Given that the 
$z = 0.6846$ absorption system is believed to originate in a late-type spiral 
disk \citep{lehar2000}, we will, in the absence of additional information, 
assume $T_x = 10$~K, a typical temperature in dark clouds in the Milky Way, 
to estimate the OH column density. This yields $\noh = 2.3 \pm 0.06 \times 
10^{15} ({T_x / 10}) ({0.4 / f})$~\cm; note that this is slightly different 
from the value $\noh = 2.65 \times 10^{15} ({T_x / 10}) ({0.4 / f})$~\cm, 
obtained by Kanekar \& Chengalur (2002) from the lower resolution ($\sim 9.4$~\kms) 
spectrum. The present value is obtained from the 1~MHz bandwidth, $\sim 
2.4$~\kms~resolution spectrum.

The OH column density of the absorber can be used to estimate the HCO$^+$ and 
H$_2$ column densities, by the relations $N_{\rm HCO^+} \approx 0.03 \times N_{\rm OH} $ 
and $N_{\rm H_2} \approx 1 \times 10^{7} \times N_{\rm OH}$ (\citealt{liszt99}; see 
also \citealt{kanekar2002}). This yields $N_{\rm HCO^+} = 6.8 \times 10^{13}$~\cm, 
in good agreement with that measured from the HCO$^+$ line ($N_{\rm HCO^+} = 
7.4 \times 10^{13}$~\cm; \citealt{wiklind95}). We also obtain $N_{\rm H_2} = 2.3
\times 10^{22}$, which is, interestingly enough, in excellent agreement with the 
value of Gerin et al. (1997) ($N_{\rm H_2} = 2 \times 10^{22}$~\cm), using the 
${}^{17}{\rm CO}$ line, but a factor of 20 smaller than the estimate of $N_{\rm H_2} = 
5 \times 10^{23}$~\cm, from the CO absorption \citep{wiklind95}. 

\begin{table}
\label{table:fit}
\begin{center}
\caption{Three-component Gaussian fit to the HI absorption}
\vskip 0.1in
\begin{tabular}{@{}|c|c|c|c|}
\hline
&&& \\
Component & 1 & 2 & 3 \\
&&& \\
\hline
&&& \\
Frequency (MHz)     & 843.136        & 843.216 & 843.126 \\
Redshift, 	    & 0.684669(5)    & 0.684510(7)   & 0.68469(4) \\
Line flux (mJy)     & $63.3 \pm 6.6$ & $39.6\pm 6.3 $& $13.5 \pm 5.6$    \\
FWHM (\kms)         & $26.6\pm 4.2$  & $19.8 \pm 4.6$  & $115 \pm 30$   \\
&&& \\
\hline
\end{tabular}
\end{center}
\end{table}

We next attempt to decompose the HI absorption into its components by 
simultaneously fitting Gaussians to the three main absorption features; 
we do not fit to the OH profiles as the signal-to-noise ratio of the wide 
component in the high resolution OH spectra is too low to get a stable fit. 
While it is quite unlikely that the net result of absorption against 
the components of the ring is indeed a Gaussian, the decomposition will be used 
solely to quantify the velocity spread of the absorption profile, in order to 
estimate the rotation velocity of the absorbing galaxy. We note that the 
results do not change significantly if we assume that the broad feature has a 
``Top-hat''-like shape (while retaining a Gaussian shape for the two deep 
components). 

Figure~\ref{fig:spectra}[A] shows the 3-Gaussian fit to the Hanning smoothed 
HI absorption profile; here, the Hanning smoothed (and re-sampled) 
7~\kms~resolution spectrum is plotted as solid squares while the 3-Gaussian fit is shown 
as a solid line. Attempts were also made to fit only two Gaussians to the 
profile. The dashed line in fig.~\ref{fig:spectra}[A] shows the best 2-Gaussian fit;
this fails to reproduce the wide absorption trough on either side of the 
two main components. Two-component fits were thus found to leave clear residuals, 
indicating that a 3-component fit was indeed necessary. The parameters of the 
3-Gaussian fit are listed in Table~1; it should be pointed out that the fit 
to the broad absorption has only a weak ($\sim 2.4\sigma$) significance. Deeper 
HI observations would be useful to test the reality of this feature, which would 
be very interesting if it were shown to indeed arise against the radio ring. 
Since the ring is only prominent at low frequencies (it is almost undetectable at
at 22~GHz but has more flux than Image~B in the 1.67~GHz EVN observations of 
Patnaik et al. 2003), deeper HI observations would provide the best sampling 
of the large scale kinematics of the lens galaxy. Finally, the FWHM of the broad 
absorption component is $115 \pm 30$~\kms; more 
relevant, the spread between points on the spectrum at which this component falls 
below the $1 \sigma$ level is $\sim 140$~\kms. This gives the total spread of 
the absorption against the radio ring. 

The 5~GHz VLBI image of Patnaik et al. (1993) (see also \citealt{biggs2001}) shows 
that image B lies roughly at the centre of the ring while the two components of image A 
lie outside the ring, at a distance of $\sim 200$~mas. The fact that the narrow absorption 
features due to the components of image A lie quite close to the centre of the broad absorption 
trough implies that these components lie close to the kinematic minor axis of the absorbing galaxy; 
it is thus difficult to draw strong conclusions about the rotation curve of the galaxy 
from these absorption lines. The observed lack of absorption against component B makes 
it likely there is a ``hole'' in the HI distribution near the centre of the $z = 0.6846$ 
absorber, similar to the situation in the Milky Way and other local spirals. As mentioned 
above, the wide HI absorption is likely to stem from absorption toward source 
components in the Einstein ring. This implies that the line-of-sight rotation velocity 
at 0.18$''$ from the lens centre (i.e. at the location of the source components in the 
ring) is approximately 70~\kms. Since the lens is at an inclination of around 25 degrees, 
the rotation velocity is $\sim 150$~\kms, fairly reasonable for an ordinary spiral 
galaxy.

In conclusion, we have detected both the 1665~MHz and 1667~MHz OH transitions in 
absorption in the $z = 0.68468$ gravitational lens toward B0218+357 and have also 
found a wide absorption component in the HI profile (at low significance). The 
redshifts of peak OH absorption are in good agreement with that of the HI, as well 
as with those of molecular lines discovered earlier in the absorber. The HI 
absorption is spread over 140~\kms~while both OH lines have a velocity spread of 
$\sim 100$~\kms. The OH column density of the absorber is $\noh = 2.3 \times 10^{15} 
({T_x / 10}) ({0.4 / f})$~\cm. Finally, we estimate that the rotation velocity of the 
galaxy is $\sim 150$~\kms, at around 1.5~kpc from the centre.

\vskip 0.15in
\noi {\bf Acknowledgments}

\noi We thank the staff of the GMRT that made these observations possible. GMRT is run by 
the National Centre for Radio Astrophysics of the Tata Institute of Fundamental Research.
The Westerbork Synthesis Radio Telescope is operated by the ASTRON (Netherlands Foundation 
for Research in Astronomy) with support from the Netherlands Foundation for Scientific 
Research (NWO).

\end{document}